\newcommand{\red}[1]
{{\leavevmode\color{black}#1}}
\newcommand{\blue}[1]
{{\leavevmode\color{black}#1}}

\documentclass[sigconf,nonacm]{acmart}
\AtBeginDocument{%
  \providecommand\BibTeX{{%
    \normalfont B\kern-0.5em{\scshape i\kern-0.25em b}\kern-0.8em\TeX}}}

\acmSubmissionID{5230}

\begin{document}

\title [On the Go with AR] {On the Go with AR: Attention to Virtual and Physical Targets while Varying Augmentation Density}

\author{You-Jin Kim}
\email{yujnkm@tamu.edu}
\orcid{0000-0003-0903-8999}
\affiliation{%
  \institution{Texas A\&M University,\linebreak College Station, USA}
  \country{}
}

\author{Radha Kumaran}
\email{rkumaran@ucsb.edu}
\affiliation{%
  \institution{University of California,\linebreak Santa Barbara, USA}
  \country{}
}

\author{Jingjing Luo}
\email{jingjingluo@ucsb.edu}
\affiliation{%
  \institution{University of California,\linebreak Santa Barbara, USA}
  \country{}
}

\author{Tom Bullock}
\email{tombullock@ucsb.edu}
\affiliation{%
  \institution{University of California,\linebreak Santa Barbara, USA}
  \country{}
}

\author{Barry Giesbrecht}
\email{giesbrecht@ucsb.edu}
\affiliation{%
  \institution{University of California,\linebreak Santa Barbara, USA}
  \country{}
}

\author{Tobias Höllerer}
\email{holl@cs.ucsb.edu}
\affiliation{%
  \institution{University of California,\linebreak Santa Barbara, USA}
  \country{}
}

\renewcommand{\shortauthors}{Kim et al.}

\begin{abstract}
Augmented reality is projected to be a primary mode of information consumption on the go, seamlessly integrating virtual content into the \red{physical} world. However, the potential perceptual demands of viewing virtual annotations while \red{navigating} a physical environment could impact user efficacy and safety, and the implications of these demands \red{are not well understood}. Here, we investigate the impact of virtual path guidance and augmentation density \red{(visual clutter)} on search performance and memory. \red{Participants walked along a predefined path, searching for physical or virtual items. They experienced two levels of augmentation density, and either walked freely or with enforced speed and path guidance.} Augmentation density impacted behavior and reduced awareness of uncommon objects in the environment. \red{Analysis of search task performance and post-experiment item recall revealed} differing attention to physical and virtual objects. \red{On the basis of these findings we}  outline considerations for AR apps designed for use on-the-go.

\smallskip
\textit{ This is a preprint version of this article. The final version of this paper can be found in the Proceedings of ACM CHI 2025. For citation, please refer to the published version. This work was initially made available on the author's personal website [yujnkm.com] in March 2025, and was subsequently uploaded to arXiv for broader accessibility. }

\smallskip

\textit{ This paper is based on the final project for You-Jin Kim's PhD dissertation, completed while the author was at the University of California, Santa Barbara (UCSB), where studies were carried out. The first author, You-Jin Kim, wishes to express his sincere gratitude to the vibrant communities at UCSB. Their friendship, encouragement, and support were invaluable throughout his six-year PhD journey. }
\end{abstract}



\begin{teaserfigure}
  \includegraphics[width=\textwidth]{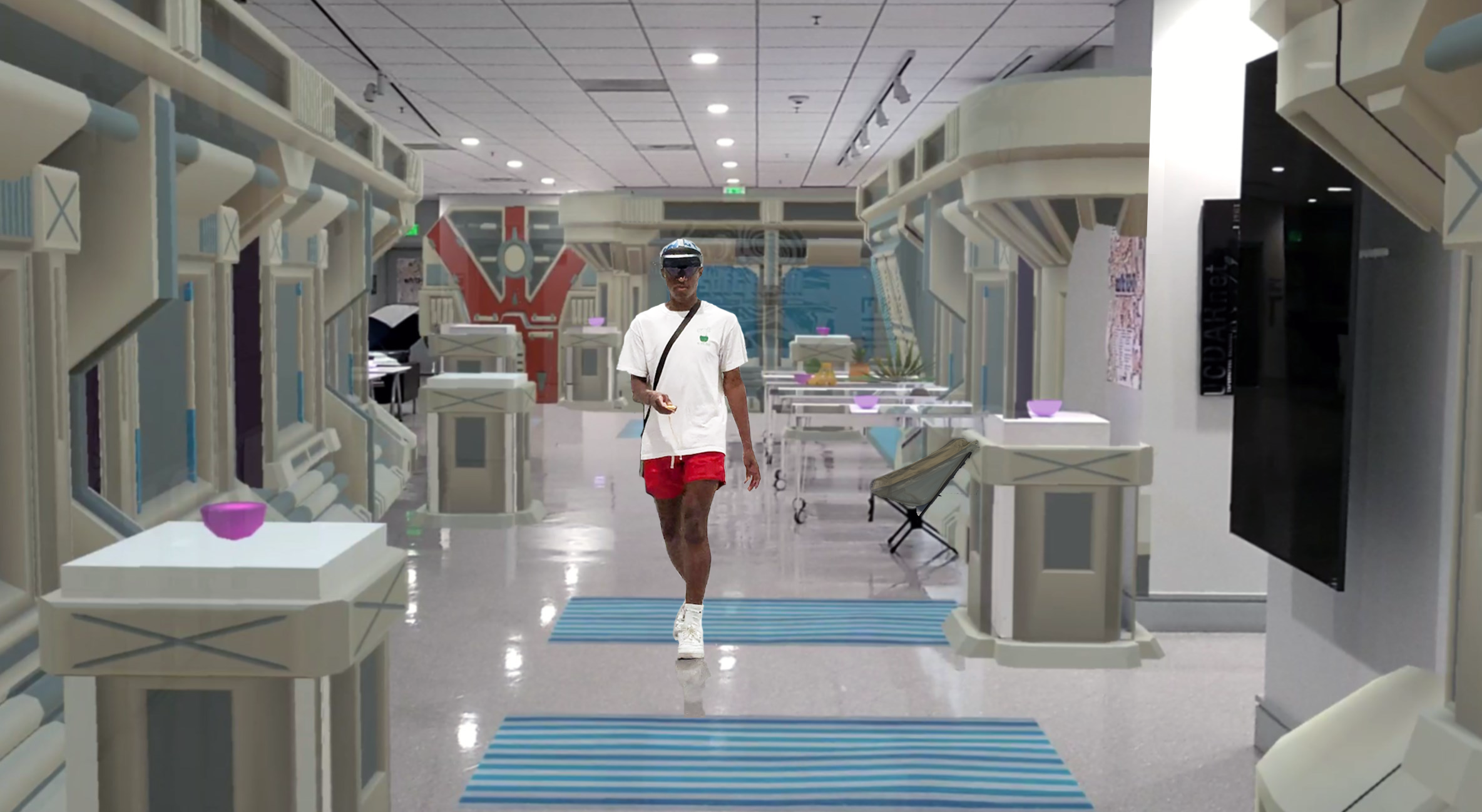}
  \caption{A participant searching for virtual and physical targets in our augmented environment. \red{In this example an image of the participant is superimposed onto a mixed-reality capture of the scene for illustration purposes.}}
  \Description{A user searching for virtual and physical targets in the AR Environment.}
  \label{fig:teaser}
\end{teaserfigure}


\maketitle

\section{Introduction}

Augmented Reality (AR) headsets, such as the Microsoft HoloLens 2, Magic Leap 2 and Xreal Ultra blend virtual content seamlessly with the physical world using sensors and advanced optics and computer vision. AR is projected to be a primary mode of information consumption in the future, thanks to its ability to integrate virtual content into the user's physical surroundings. These unique capabilities of AR could be particularly useful for  information consumption on the go, when more powerful but bulkier devices such as PCs, laptops, or tablets are not convenient~\cite{hollerer2004mobile}. 

AR has demonstrated advantages in supporting search and navigation~\cite{bauerfeind2021navigating}, and could prove useful in scenarios such as emergency search and rescue~\cite{zhu2021virtual, luksas2022search}, exploratory learning~\cite{ibrahim2018arbis, garzon2019meta} or tourism~\cite{egger2020augmented}. \red{Always-on AR smart glasses are also expected to be widely used for personal everyday use in the near future, projecting virtual content into the physical world as users navigate larger environments. Given that such AR use would make information more conveniently accessible than through smartphone use while walking, which has its own increasingly demonstrated cognitive costs~\cite{schildbach2010investigating,lin2017impact,mourra2020using,laurier2017mediated}, we need to understand the potential impact of AR consumption on walking users' behavior and cognitive processes before designing head-worn AR applications that could support such tasks as navigation and wayfinding in unfamiliar environments, or  applications such as entertainment, education, communication, corporate duties, or personal information management during common daily walks~\cite{chang2024experience}. }

The introduction of virtual content into environments that are already crowded could lead to cognitive or perceptual overload, potentially distracting from the user's primary tasks and even jeopardizing their safety by reducing their awareness of the physical surroundings. \red{There is a limited understanding of the cognitive phenomena that are important, and most impacted, when users perform AR tasks while walking to navigate an environment.} It is therefore crucial that we better understand user responses to increasingly dense augmented environments. Accordingly, the first goal of this paper is to investigate the impact of different levels of environment augmentation density on user performance in a mobile AR search task. \red{Environment search is a suitable generic but representative task for several application scenarios that require some attention to the physical world while benefiting from virtual augmentations. }

The main strength of AR is its ability to situate virtual content in the real world, with the intention of creating a blended space that users perceive as a single environment. However, even with the current state-of-the-art technology, there are still visible differences between the real environment and renderings of virtual content. Significant technological advances are still required before the distinctions between virtual and physical worlds can be eliminated, and indiscernibility of real and virtual items  may not even be desirable in all applications. In AR-guided search tasks, users must simultaneously focus on both real and virtual environments, so it becomes essential to understand how users attend to and interact with these blended spaces. Therefore, the second goal of this paper is to better understand how users interact with goal-relevant objects in both the virtual and physical worlds, by having them search for both physical and virtual items.

The use of AR in real-world environments has been shown to impair user safety, due to the potential of virtual content obscuring real-world objects, and reduced situational awareness~\cite{krupenia2006does, kumaran2023impact}. Future AR on-the-go applications are likely to be used in much less controlled environments than current experiments, \red{and may also display various notifications while in use, potentially distracting users as they navigate their environment}. Users may want to use AR on a variety of both destination-driven and more meandering walks. AR may even be the interface of choice guiding a user along a path, similar to how Live View on Google Maps provides navigational guidance on AR-enabled smartphones~\cite{GoogleLiveView}. Whatever AR guidance is provided, be it for navigation or exploration purposes, ensuring the walker's safety and the safety of others in the vicinity will be essential. 

\red{Always-on AR annotations or guidance mechanisms may increase cognitive and perceptual demands on the user, requiring them to dual-task and potentially reducing their situational awareness of the physical world, hence the impact of these mechanisms on behavior must be studied. }Our third goal is to investigate the impact of a visual guide that controls the user's walking path while they perform a task that requires scanning the environment for relevant items.

Here we designed a search and classification task in which participants searched the environment for semi-spherical "gems", and classified gems into two categories based on the presence (or absence) of surface markings on the gem that required closer inspection to discern. To address our first goal, we manipulated augmentation density in the environment, such that participants performed the task in both lightly augmented and heavily augmented environments. To address our second goal, we had participants search for both physical and virtual gems, which were matched for perceptual characteristics as closely as possible. Our study is the first to include both real and virtual targets in such a search task, to the best of our knowledge. To address our third goal, in half the experimental trials we introduced a "spotlight" ring annotation at approximately the user's waist height that moved along a predefined path through the environment. Participants were required to stay within the bounds of the ring when it was present. 

\vspace{0.5in}

Our key findings suggest that
\begin{itemize}
    \item environmental clutter significantly impacts user perception and task performance, as evidenced by 
    \begin{itemize}
         \item increased head rotation in the presence of high clutter and 
         \item reduced awareness of a highly conspicuous non-target object (a model of Godzilla). 
    \end{itemize}
    \item There appears to be a divide in attention between the physical and virtual scenes, with physical targets being more easily detected.
    \item In contrast to prior work that reported for AR search tasks involving purely virtual targets that physical scene objects were far less memorable in a recall task than virtual objects, a more balanced recall performance among virtual and physical objects in this task provides new evidence regarding the possible causes of this effect.
\end{itemize}
  
Together, these results lead to valuable insights for the future development of AR experiences for walking users.

\section{Related Work}
\label{sec:related-work}

Our present study is inspired and informed by three broad areas of relevant literature. In this section, we summarize key findings from studies that explore perception and attention in AR, the blending of physical and virtual worlds in AR, and search for goal-relevant items in AR.

\subsection{Perception and Attention in Mixed Reality Environments}

In recent years, standalone AR headsets have become more powerful and capable, enabling them to overlay increasingly rich and complex layers of augmentation into the physical environment. However, the current technological constraints of AR headsets prevent them from projecting truly photo-realistic virtual information into the physical world, meaning that the user can always still clearly differentiate between real and virtual, thus preventing seamless blending of the two worlds. Optical see-through head-mounted displays (OST-HMD), such as the Hololens 2 used in this study, have a restricted field of view~\cite{ren2016evaluating}, poor color resolution~\cite{Livingston2013Basic, Renner2017Attention}, do not effectively render black or dark annotations without carefully crafted halo or highlighting techniques \cite{manabe2019shadow} and cannot function in brightly lit environments due to their use of an additive light model~\cite{Erickson2020Exploring}. These lighting and display limitations can induce eye fatigue \cite{Livingston2013Basic, Gabbard2019Effects, Paul1960Membrane} and can impact color perception \cite{Gabbard2010More, Hincapie-Ramos2014SmartColor} and depth perception~\cite{Swan2007Egocentric, Smith2015Optical, Adams2022Depth} in the user. Fusion of virtual information with the physical world, especially when the virtual information is perceptually inaccurate or degraded, raises concerns about the potential impact of AR technologies on how humans perceive and attend to information in both worlds. \red{To better understand these potential impacts, we draw upon modern theories of attention, visual search, and investigations into attention within mixed reality environments. Given our paper's central theme, we narrow our discussion to selective attention and visual search. Recent study~\cite{chang2024experience}, which explores AR interfaces in real-world walking scenarios, provides valuable insights, emphasizing the importance of our research into how AR influences perception and attention.}

Search of a natural visual scene requires that we deploy selective attention \cite{Cohen2011Natural}, because we cannot process everything in the scene all at once. Contemporary models of selective attention assert that selection can be guided by a wide range of different factors including but not limited to current goals, physical salience of items in the scene and prior search history \cite{awhTopdownBottomupAttentional2012a,shomsteinAttentionPlatypuses2023,Wolfe2017Five}. In other words, attention can be deployed voluntarily across a room to scan for something specific like a set of keys; captured involuntarily by a sudden movement; and drawn towards items associated with a positive outcome. The perceptual difficulty of a primary task can also determine whether an individual will become aware of task-irrelevant items in a visual scene, with higher task demands leading to reduced competition from competing distractor items \cite{Lavie1995,Lavie1997}. Furthermore, in the absence of focused attention we can fail to detect highly salient and unexpected objects that appear in our environment, as demonstrated in the well-known "Invisible Gorilla" study \cite{Simons1999}. Together, these models and examples from traditional lab-based cognitive studies confirm that attention is highly susceptible to a range of different goal and environment related influences.  

Both voluntary and involuntary modes of attentional control can potentially be disrupted when a user is placed in a mixed reality environment. For example, in dense augmented environments, newly entering goal-relevant target items can be missed due to occlusion by virtual “clutter” in the scene \cite{ellis1998localization, kim2024audience, trepkowski2021multisensory}, and this clutter can impair user interactions with both physical and virtual objects \cite{satkowski2021investigating, chiossi2024understanding}. Furthermore, when compared with user experiences in physical environments, AR can have a negative impact on users’ attention \cite{Biermeier2024Measuring, Eyraud2015Allocation}, particularly during the decision-making phase of a driving task \cite{Eyraud2015Allocation}. Studies on the use of mixed reality in an outdoor wide-area search task also indicate that attention to virtual objects may reduce situational awareness of objects in the physical environment \cite{kim2022investigating,kumaran2023impact} (see Section 2.3 for discussion of wide-area search).

In summary, this related work suggests that AR can have a profound impact on user's attention. The three main goals of this study are designed to assess aspects of AR that may impact user attention: augmentation density, search for both physical and virtual task-relevant items in the environment (see next section), and the additional dual-tasking demands of navigational guidance on attention.

\subsection{Physical and Virtual Objects in AR Experiences}

As Mixed Reality (MR) technologies become more advanced (e.g. higher resolution, computing power), tangible user interface systems, which integrate virtual content into physical worlds \cite{Kato2002Virtual}, such as ARTable \cite{Park2006ARTable}, have blurred the lines between the physical and virtual. In Cross-Reality (CR) environments, transitional navigational techniques such as MagicBook \cite{Billinghurst2001MagicBook}, which transport users into the virtual environment of book pages, and the creation of virtual environments from live video feed of the physical environment \cite{Tatzgern2014Transitional} are examples of how there is increased interest in blending physical and virtual worlds and transitioning between them. Studies on this relationship have measured traditional approaches like “step into” or “reach into” as well as newer methods like the “blended space,” where users can see the source and destination environment simultaneously \cite{Cools2022Blending}.  

Mixed reality environments have been explored in various studies and projects \cite{moacdieh2012eye, thomas2002first, kumaran2023impact, raskar1998office}. Notable examples of these enhanced experiences include projects such as Remixed Reality, Reality Distortion Room, and Room2Room \cite{kim2023reality, Lindlbauer2018Remixed, pejsa2016room2room}. These projects utilize the physical layout of a room and employ spatial augmented reality to extend user interaction and experience within these blended spaces, where interaction is expanded through imagination. Additionally, the Blended Whiteboard project \cite{gronbaek2024blended} suggests interacting with a physical object, in this case, a whiteboard, from each user's perspective, allowing them to experience seamless interaction as if they are in the same location, even when they are not.

Assessing the relationship between physical and virtual objects, studies have examined the ways virtual objects adapt to shifting physical environments, such as those modeling illusionary episodes \cite{cheng2021semanticadapt, kari2023scene, Tatzgern2014Transitional}. Separate from the environment, virtual objects can also be manipulated by going through recoloring, teleportation, moving, copying, scaling in real time, and customization such as in MineXR \cite{Lindlbauer2018Remixed, cho2024minexr}.

With increasingly complex blended spaces, virtual augmentations such as AR cues and adaptive clustering have been implemented to direct user attention to target objects and simplify distracting environments respectively \cite{tatzgern2016adaptive, raikwar2024beyond}. AR cues given from a first-person view (FPV) have particularly shown increased ease in following instructions and performing tasks \cite{lee2020enhancing, kim2024spatial, lilija2020put}. Direct manipulation of objects also allows users to better navigate spatial recordings of AR environments.

Using current augmented reality (AR) technologies, users perceive two worlds due to incongruencies of display, rendering, blend between physical and virtual content, and the ability of AR to modify and create environments that do not exist in the real world \cite{Westermeier2023Exploring, Westermeier2023Virtualized, Lindlbauer2018Remixed, cheng2019vroamer}. In the real world, people experience source confusion, which is when memories are attributed to the wrong source (e.g. confusing pictures seen with real experiences) \cite{bonnail2024real}. Since virtual objects have greater media richness than physical objects, researchers have speculated that VR could provide more memory suggestions in ways that are autobiographical (real life) rather than episodic (shallow, isolated events) \cite{segovia2009virtually, schone2019experiences, bonnail2024real}. 

Previous work has shown that the phenomenon of attentional tunneling, which is where users focus more on virtual objects rather than physical objects \cite{wickens2009attentional}, is affected by the number of virtual tasks \cite{syiem2021impact}, accurate cues to complete tasks \cite{yeh2001display}, and level of engagement the tasks require \cite{dixon2014inattentional}. 

As the second goal of our study, we explore the phenomena of attention tunneling and source confusion during natural walking while exploring an AR environment. We compare attention to physical and virtual target objects during a search task, as well as investigating attention tunnelling and source confusion through a memory task involving task-goal irrelevant objects in the environment (which participants are only informed of after the experiment).

\subsection{Search and Mobile Information Browsing in Augmented Reality}

AR's ability to overlay digital information into the physical world unlocks many potential use cases for assisting with search tasks during various different scenarios. For example, emergency search and rescue, using always-on AR to explore a new place, or locating a target in a crowded space. For this reason, search has been studied in mixed reality environments, with two broad types of search tasks often used - informed search tasks, where participants are guided to the location(s) of the target(s) using virtual cues~\cite{rompapas2019towards, kruijff2018influence, kishishita2014analysing, biocca2006attention, kim2023reality}, or uninformed search tasks, where participants are not given any information about the location of the target(s)~\cite{lee2020effects}. We note that in the present study participants were not cued to the locations of the targets, because the focus of the experiment was to observe natural attention patterns in the different experimental conditions.

Much of the literature on search in XR does not involve any natural locomotion, likely due to technological limitations. Given the portability of AR and enormous future potential for it to be used on the go, the lack of locomotive AR studies represents a critical gap in the literature. \red{Initial studies have explored mobile information browsing via window-based AR workspaces that are either oriented with walls in the environment or following the user in various ways \cite{lages2019walking,chang2024experience}, but the impact of 3D AR scene density has not yet been formally explored.} Recent advances have enabled studies on users navigating and searching larger real-world environments while wearing mixed reality headsets, scouting for virtual target items under different lighting conditions ~\cite{kim2022investigating} and with the assistance of different navigational aids ~\cite{kumaran2023impact}.  The present study makes several important novel contributions to our understanding of locomotion-based search in AR, by assessing the impact of augmentation density, physical and virtual items and path guidance in a locomotion-based search task.

\section{Experimental Design}

In the present study, participants performed a search task, where they searched for 12 target gems in an indoor area (208.54 m2, or 2,244 ft2) augmented with virtual and physical furniture, wall augmentations and other virtual objects.

\begin{figure}[t]
\centering \includegraphics[width=\columnwidth]{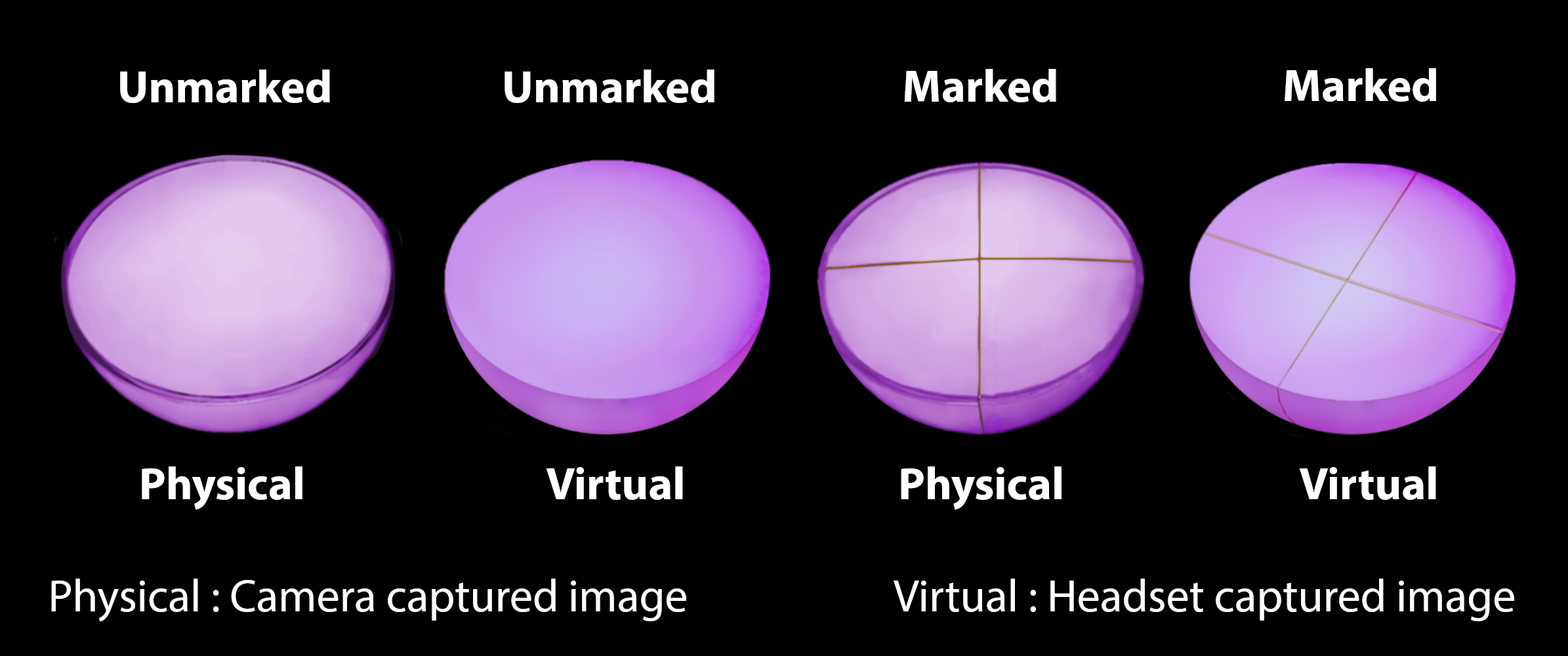}
 \caption{Four classes of gems encountered by participants. From left to right: Unmarked-Physical, Unmarked-Virtual, Marked-Physical, Marked-Virtual. All gems emit purple light, and the virtual gems were designed to closely resemble the physical gems.}
 \label{fig:gems}
 \end{figure}

\subsection{Task}

Participants performed two tasks in the experiment - a search and classification task (8 trials total) and a surprise object recall task designed to probe memory for goal-irrelevant objects in the environment, which was administered  after all the search task trials were completed.

In each trial, participants had to walk through the L-shaped environment and classify each gem they found into one of two categories - marked  or unmarked (Figure \ref{fig:gems}) - by clicking buttons on a handheld controller. If a participant classified the gem and their response was registered by the system, a small yellow sphere appeared above the gem and remained visible throughout the trial.

Participants were instructed to walk through the environment in a single direction, ensuring that they explored new areas without retracing their steps or revisiting any previously encountered sections of the environment. On trials when path guidance (via a spotlight ring annotation moving above the floor) was provided, participants were instructed to maintain a consistent position within the ring throughout the trial as they performed the task.

At the end of all trials participants answered an object memory questionnaire, where they were given a list of twelve objects (six absent and six present in the environment). They had to indicate which objects were actually present in the environment, and if present, whether they were physical or virtual.

\begin{figure}[t]
\centering \includegraphics[width=\columnwidth]{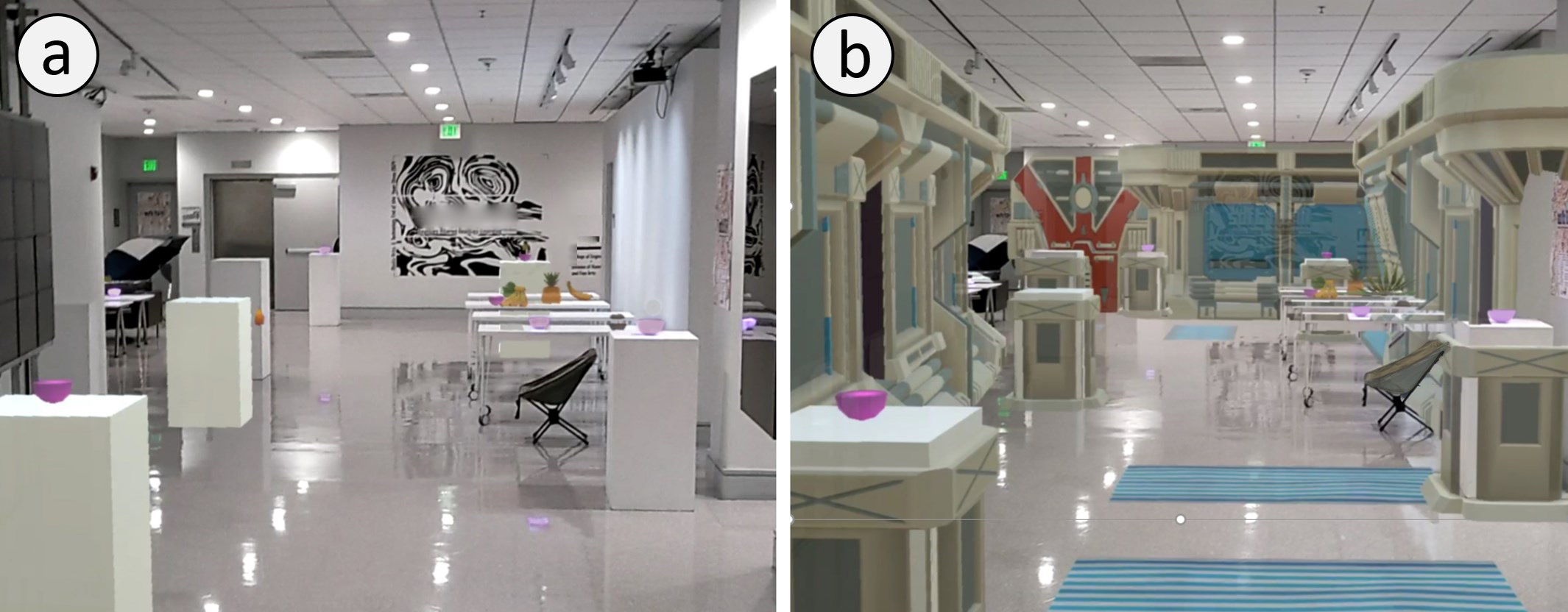}
 \caption{Participants experienced two levels of virtual augmentation (environmental clutter) in the hallway the experiment was conducted in. Images depict the environment with low (a) and high (b) levels of augmentation.}
 \label{fig:clutter}
 \end{figure}

\subsection{Design}

In keeping with our three key goals, we manipulated augmentation density, gem type and path guidance in a locomotion-based search task. Specifically:

\paragraph{Augmentation Density} Participants performed the search task under two different augmentation levels: low augmentation (Figure \ref{fig:clutter}a, physical environment with few virtual additions) and high augmentation (Figure \ref{fig:clutter}b, approximately equal amounts of physical and virtual content). \red{For the purposes of this study, physical content was defined to include both objects in the environment, as well as physical structures such as walls and pillars. The ceiling and floor, which were intentionally excluded from full augmentation, were not considered. In the low augmentation scenario, no walls or objects underwent skin wrap alterations that would change their appearance. Conversely, in the high augmentation scenario, approximately two-thirds (66\%) of the wall surfaces were augmented throughout the environment. Physical objects, such as stools, were given virtual skin wraps to enhance their visual appeal, while still retaining their functionality and visibility with amplified virtual surfaces. This was done intentionally to create a distinct difference in the appearance of the two conditions. As the future of AR leans towards more extensive augmentation, we wanted to explore the effects of a highly augmented, nearly virtual environment where the physical layout and content is still preserved.}

\paragraph{Gem Type} In each trial, participants searched for six virtual and six physical gems scattered throughout the environment, which were matched as closely as possible in terms of their physical characteristics (Figure \ref{fig:gems}). Participants were not explicitly asked to detect whether the gems were physical or virtual, but gem type (virtuality) was included as a factor in our analyses. 

\begin{figure}[t]
\centering \includegraphics[width=\columnwidth]{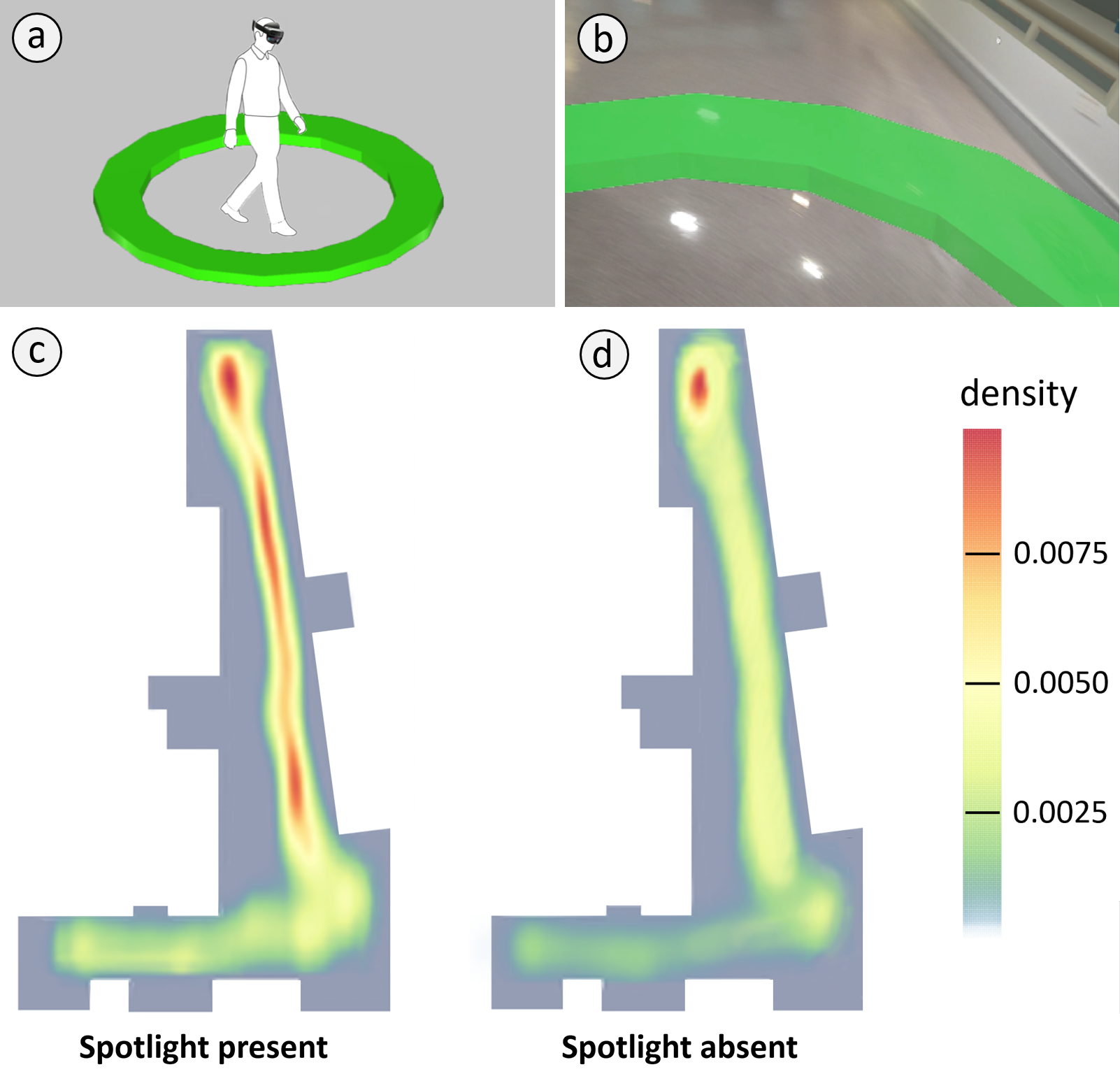}
 \caption{Environment navigation. (a) A depiction of a participant walking \red{within the "spotlight" ring} - a virtual green ring that was used to guide participants' path through the study environment in the guided conditions. (b) An example of the spotlight from a participant's perspective (captured directly from the headset). (c,d) Heatmaps depicting the locomotive activity of all participants during spotlight-present (c) and spotlight absent (d) trials. Higher temperatures indicate increased dwell times. }
 \label{fig:spotlight}
 \end{figure}

\paragraph{Path Guidance} In half of the trials, participants were guided on a set trajectory determined by the movement of a green ring ("spotlight") that moved at a steady pace of 0.92 m/s at 0.75m height above the floor from the entrance of the corridor to the back exit, taking 60 seconds in total to complete the journey (Figure \ref{fig:spotlight}a, \ref{fig:spotlight}b). In the other half of the trials the spotlight was absent, and participants were able to walk through the environment at their own pace, with the only constraint being that they were not permitted to retrace their steps. While effective in providing a predefined path, as shown in (Figure \ref{fig:spotlight}c), maintaining position within the spotlight placed an extra burden on the user's attention, essentially creating a dual task that served as an additional challenge. 

\begin{figure}[t]
\centering \includegraphics[width=\columnwidth]{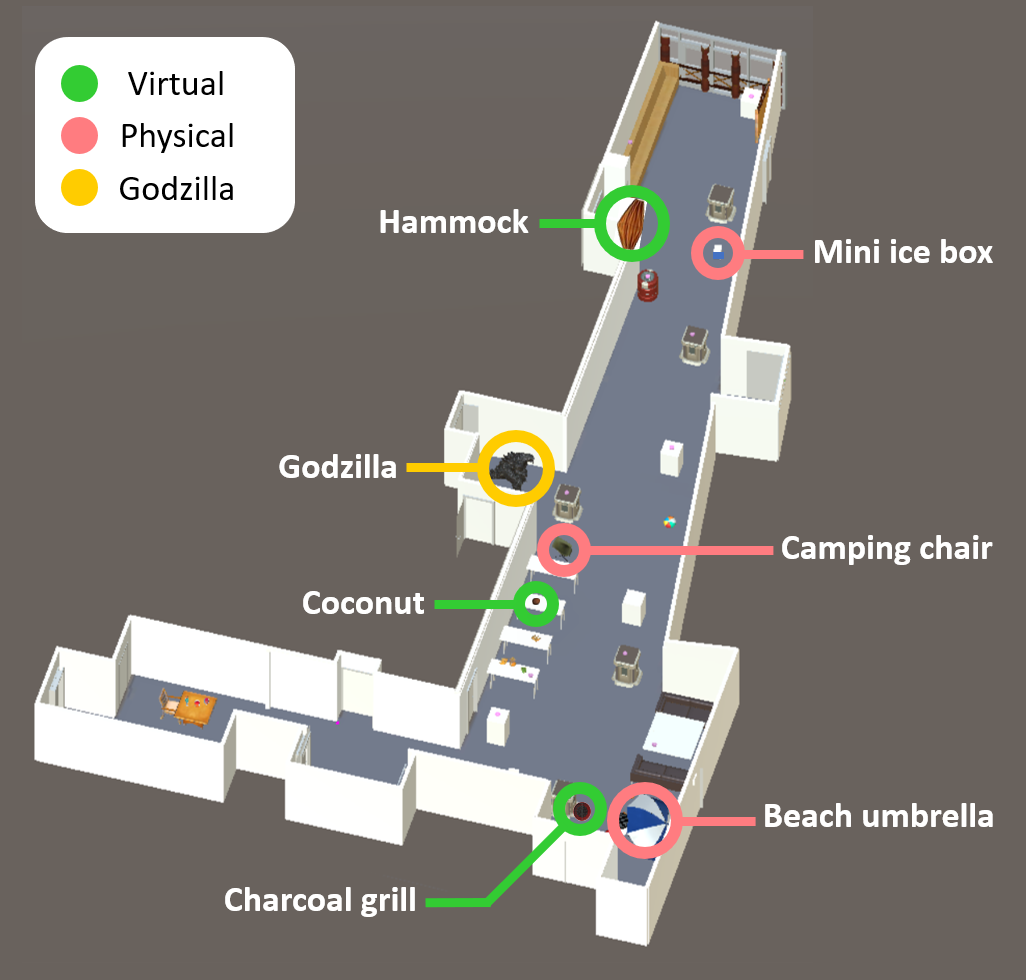}
 \caption{An aerial perspective of our user study location, with virtual objects marked in green, physical objects highlighted in pink, and Godzilla represented in yellow. From this top view, it can be observed that the distribution of both virtual and physical objects is spread out evenly across the entire location.}
 \label{fig:location}
 \end{figure}

\begin{figure}[t]
\centering \includegraphics[width=\columnwidth]{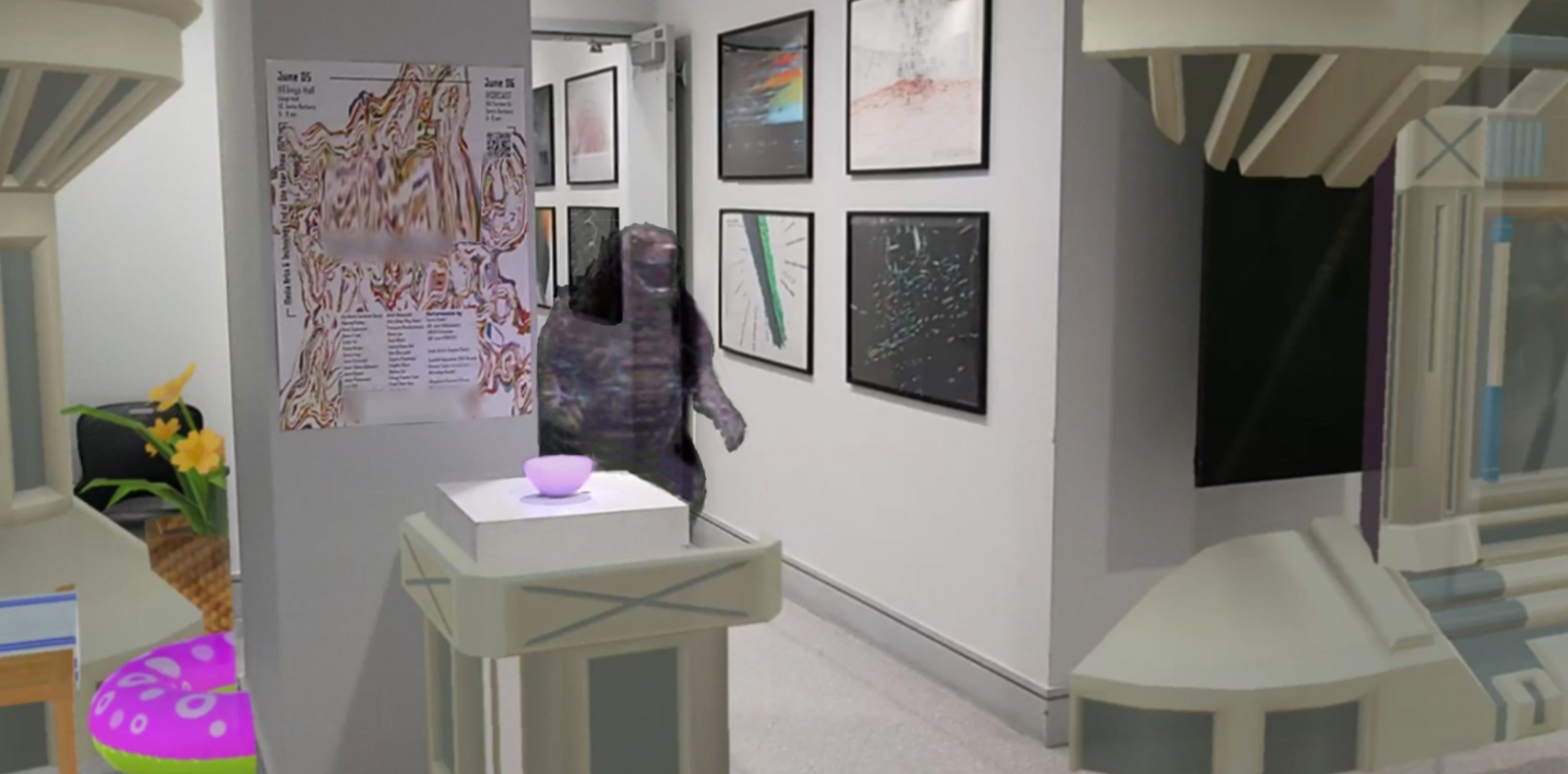}
 \caption{Highly salient object manipulation. A virtual model of the monster "Godzilla", standing 190 cm tall, was positioned approximately halfway down the hallway. Godzilla was present in every trial, and we measured whether participants perceived it by asking if they saw anything "noteworthy" at the end of each trial.}
 \label{fig:godzilla}
 \end{figure}

We also manipulated the presence of specific task-irrelevant objects in the environment, with the goal of probing participants' recall for these items once the search task was complete. Six beach-themed objects were placed throughout the environment in each trial to assess user attention and memory, including three physical ones (mini ice box, camping chair, beach umbrella) and three virtual (coconut, hammock, grill). \red{The positions of the objects were not changed between trials or conditions, and they were distributed throughout the environment as evenly as possible while still ensuring a safe walking path.} After completing the eight search trials, participants were given a surprise object recall test (see Procedure).  \red{Furthermore, we included a highly salient virtual object (a large "Godzilla" figure, standing 1.9m tall. See Figure~\ref{fig:godzilla}) in every trial, with the specific goal of probing participants' awareness for an object that was incongruent with the surroundings.} In the absence of focused attention, it is possible for individuals to fail to become aware of highly salient items in the environment - a phenomenon known as inattentional blindness \cite{Simons1999}. An additional goal of of our study was to test the impact of AR on susceptibility to inattentional blindness. Probing participants' awareness of this highly salient item at the end of each trial meant that we could assess if/when they first noticed the item, and if they did notice it, under what conditions of augmentation and path guidance. \red{Godzilla was placed in the same position during all trials and conditions, in a passageway branching off from the main environment (as seen in Figure \ref{fig:location}). Godzilla was occluded by pillars and walls from some angles, but was positioned right behind a stool that sometimes held a target, making it highly likely that Godzilla fell within the participant's field of view during the study.}

\begin{figure}[t]
\centering \includegraphics[width=\columnwidth]{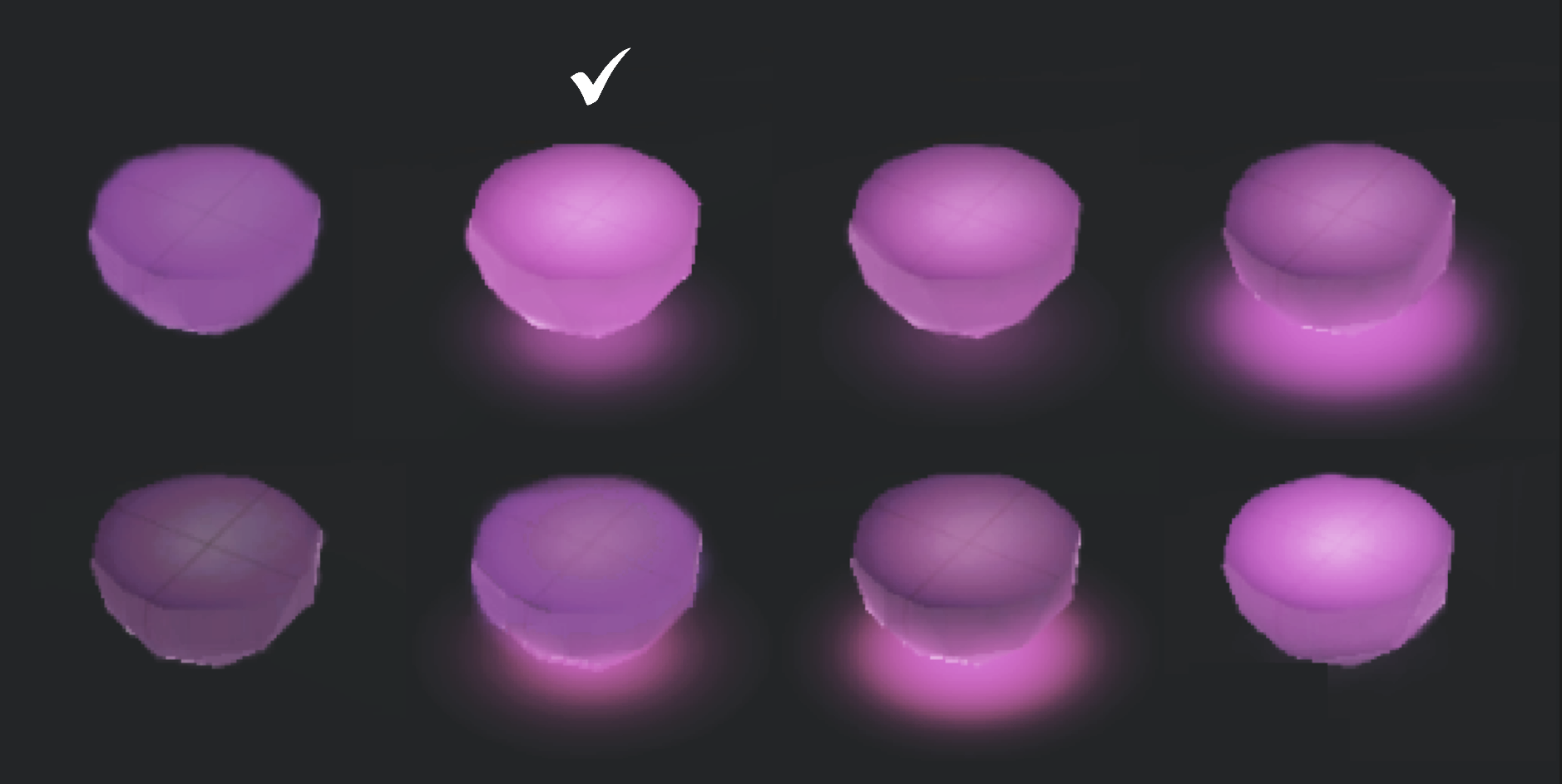}
 \caption{Twenty-five different light renderings were tested to create virtual gems that most resembled the physical gems. Eight of the twenty-five renderings are shown, with the second gem (from left ) in the top row being the most accurate and thus chosen for the study.  The image shown was captured directly from Unity, not from the headset.}
 \label{fig:gems_unity}
 \end{figure}

\begin{figure}[t]
\centering \includegraphics[width=\columnwidth]{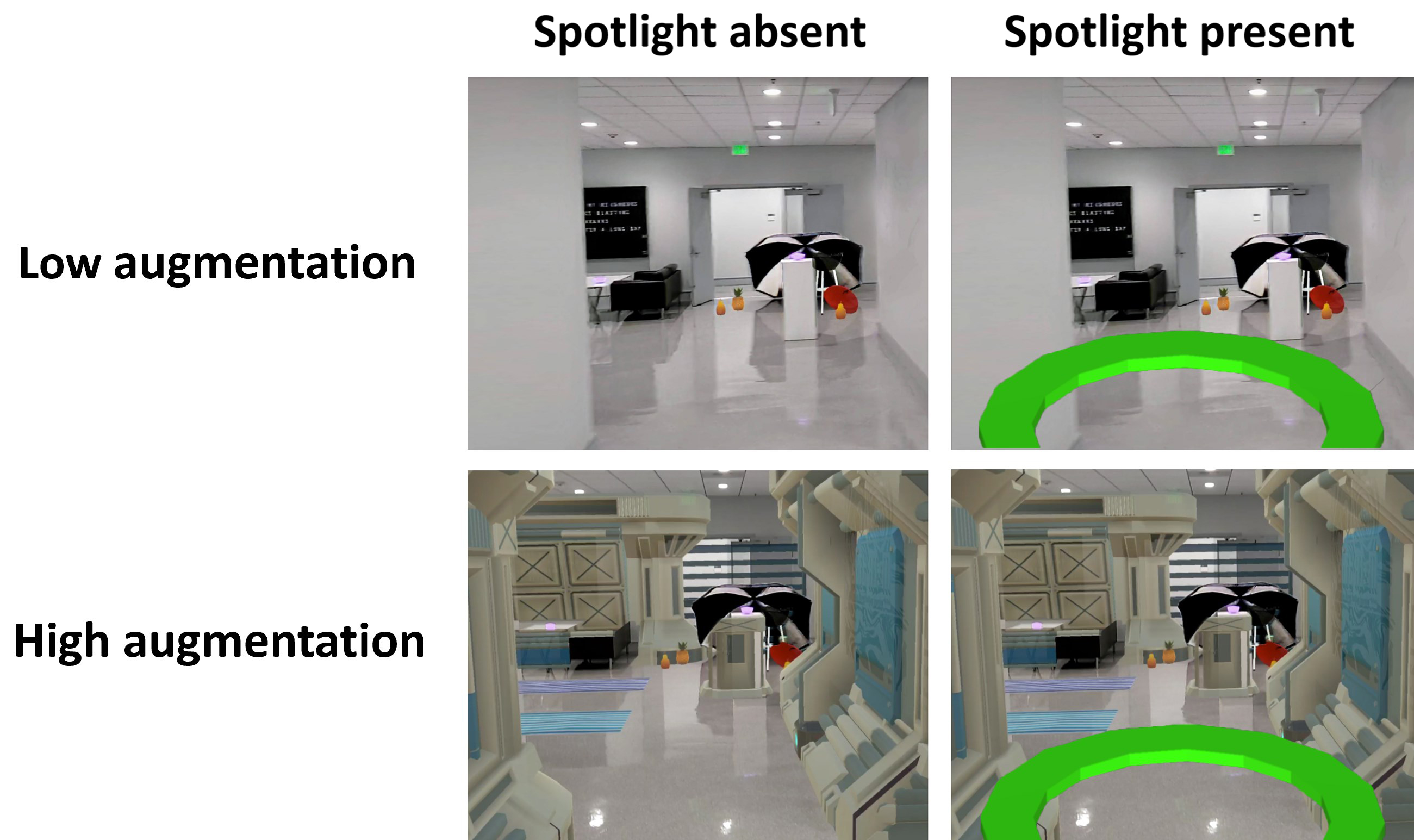}
 \caption{Four different study conditions. Examples from the full factorial combination of augmentation (low, high) and spotlight (absent, present), which creates four distinct conditions. Condition order was fully counterbalanced, and participants completed two trials of each condition, resulting in a total of eight trials for each participant. The participant's perspective at the start of a trial is depicted in each of the images.}
 \label{fig:trial-condition}
 \end{figure}

In summary, each trial was a factorial combination of augmentation density (low, high) and path guidance (spotlight ring present, spotlight ring absent) (Figure \ref{fig:trial-condition}). Participants searched for six physical and six virtual gems in each trial. In each of the eight trials participants encountered a distinct arrangement of 12 gems, with the positions randomly chosen from 21 possible locations. The eight gem arrangements (curated to ensure a balanced distribution of gems throughout the scene) were randomized across conditions for all participants using a set of Latin Square permutations. Participants completed four conditions in total, and each condition was repeated twice (eight trials total). Condition order was fully counterbalanced between participants.

\subsection{Pilot Considerations and Formative Experiments}

We conducted several pilot experiments to determine parameters of the experiment design such as the appearance of the virtual gems and pace of the gliding ring for the spotlight. 

\subsubsection{Gem Design and Rendering}
We tested various types of lights for the environment, including Point Lights, Spot Lights, Directional Lights, and Area Lights. Area Lights were the most effective due to their ability to illuminate objects from multiple directions and create softer, more subtle shading, making them ideal for blending physical and virtual gems without casting harsh shadows. To capture the subtle light bleed seen in physical gems, we used the MRTK Hover Light Shader, which mimics the effect of an area light and light bleed that may cast to the surface. The rendering of six virtual gems in this fashion did not negatively impact frame rate on the HoloLens 2.

We created 25 versions of the virtual gems and compared them with physical gems using an AR headset to determine the best match, eventually resulting in a remarkably similar appearance. Some of the candidate rendering models are shown in Figure \ref{fig:gems_unity}. The chosen design specifically matches shadow and light bleeding on the surface of the object it was mounted on, as validated in a pilot study with 5 participants.

\subsubsection{Gem Marking and Distance} 

In order to emulate the stripe markings of a physical gem as precisely as possible in our virtual gems, and to ensure that rendering closely mirrored the original gem, we developed a custom UV mapping based on a physical gem's scan. 
\red{We performed multiple tests and comparisons to achieve the best possible match, through early pilots with team members. We then conducted a pilot study with five users who did not subsequently participate in our main study. Each pilot session lasted approximately 30 minutes, and helped us select the most suitable gem rendering model from 25 pre-designed options, with each participant narrowing their choices to their top 5 and then top 2.

We then conducted tests for gem identification at distances ranging from 1 to 5 meters. During these tests, each participant was required to identify 10 virtual gems, both marked and unmarked, presented in a random sequence. Participants achieved nearly perfect scores at distances of 1 meter (100\% accuracy) and 2 meters (98\% accuracy). Accuracy was still high at a 3-meter distance (96\% accuracy), but dropped significantly beyond 4 meters (72\% accuracy), plummeting to 28\% at 5 meters. Participants noted that the images appeared pixelated at these greater distances, making it difficult to discern the presence of stripes. We therefore positioned all gems within 3 meters (at their shortest distance) of the planned walking path. During the main user study, 99.7 percent of all classifications occurred within a range of 2.6 meters.
}
\subsubsection{Gliding Spotlight}
\red{The choice of this "spotlight" path guidance mechanism, which indicates a region that the user must stay within (much like real-life spotlights in stage performances), was driven by our goal to investigate visual cues that indicate safe walking areas in an environment during a timed navigation task, akin to real-world tasks such as emergency response, evacuation training, or navigation and wayfinding. The mechanism also loosely models self-imposed casual walking speed along a familiar path, which is meaningful for future always-on AR opportunities. 

There are different ways to enforce a specific path on the user, ranging from directional arrows (that are merely suggestions of the path to follow) to a visibility bubble (that penalizes a user if they do not remain in the bubble). We chose to implement a mechanism with an intermediate level of enforcement: a moving "spotlight" ring that indicates the area to stay within with a full$\ 360^{\circ}$ visual restriction that changes color when the user is not inside (but does not penalize them in any other way). We also deliberately chose to place the visual at approximately waist height rather than on the floor, so that participants would not need to look away from the surrounding environment in order to keep track of the ring while they were actively scanning for gems. We conducted several early pilots with the ring implementation to ensure that target (and other meaningful) objects were not obstructed by the ring.}

In our pilot experiments, we tested five gliding spotlight ring speeds to determine the optimal speed for the gliding spotlight, aiming to approximate a "normal,"  non-dawdling,  purposeful walking pace. Our findings indicate that the arrived-at speed, 0.92 meters per second, ended up being only slightly higher than the median speed chosen by individuals when freely optimizing target detection accuracy and time. \red{This suggests that the differences in participant behavior between the spotlight and free walking conditions were not solely due to the time taken to complete the task.}

Although the spotlight speed is still relatively slow compared to typical \red{brisk} walking speeds, it closely matches the speed at which users walk while actively scanning the environment for targets.

\begin{figure}[t]
\centering \includegraphics[width=\columnwidth]{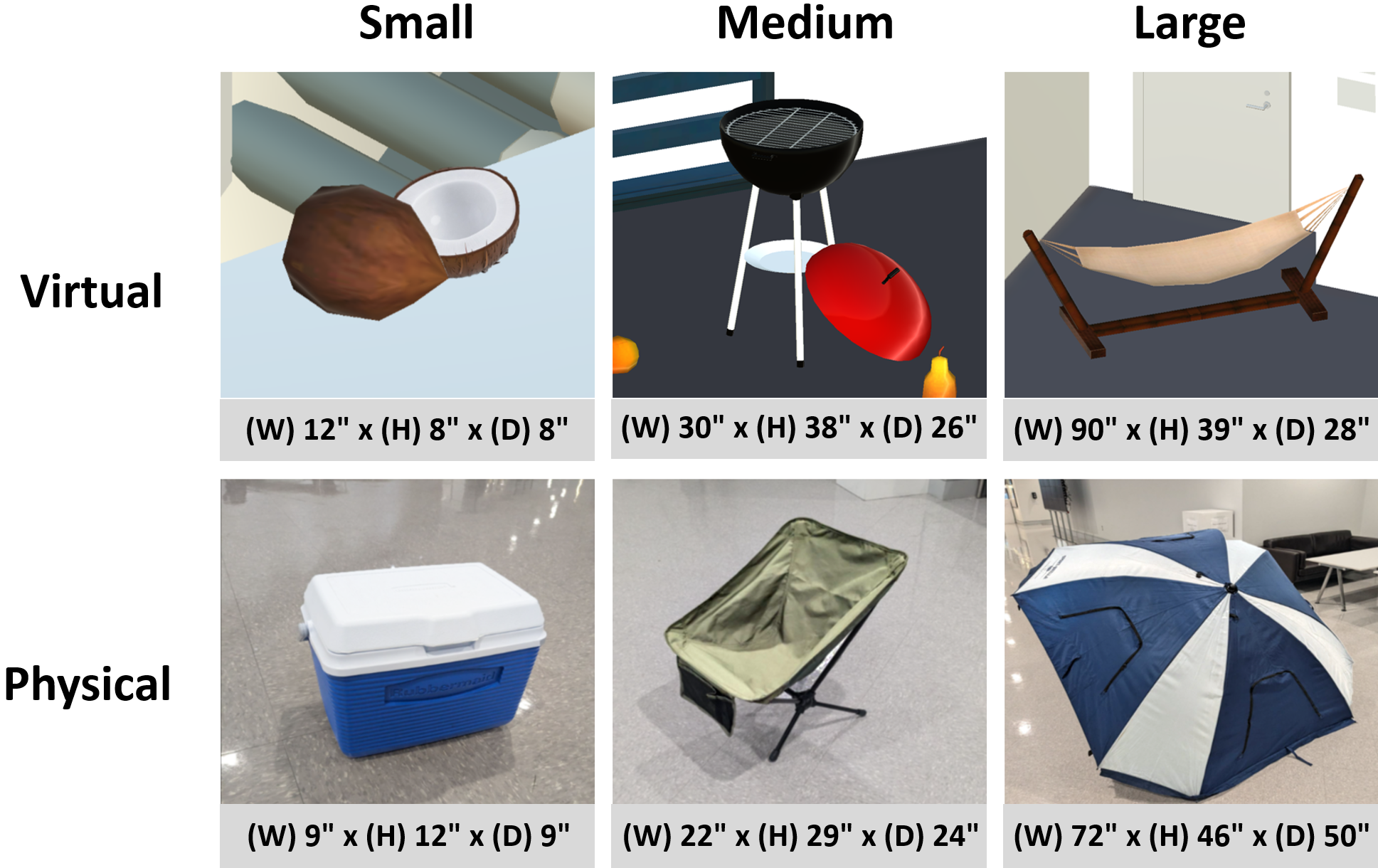}
 \caption{Additional physical and virtual objects for \red{the} surprise recall test. The environment includes six objects of different sizes that were tested on the surprise memory task. Three virtual objects (top row): a coconut, charcoal grill, and hammock. Three physical objects (bottom row): a mini ice box, camping chair, and beach umbrella. Object sizes  ranged from small to large, shown from left to right. After all eight trials of the gem search task were completed, participants were asked to recall selected objects and indicate whether each was physical or virtual. The size of each object was measured using its bounding box.}
 \label{fig:recall-objects}
 \end{figure}

\subsection{Apparatus}

The experiment was conducted on a Microsoft HoloLens 2 headset, and responses were recorded from a Bluetooth clicker, with two buttons used for the two types of gems (marked and unmarked). The physical gems were Philips Hue Go portable smart lamps, with color and brightness of all lamps in the scene controlled from the Philips Hue app. 

We measured light levels in the environment and the HoloLens-2 display brightness for a representative object from the participant’s eye position relative to the headset. Measurements were made avoiding direct light sources, looking parallel to the ground, and were collected at either ends of the hallway and at the middle. The luminance measurement for the physical environment reads (M = 336.00, SD = 105.78) while the AR objects emit (M = 487.33, SD = 166.54) lux. Measurements were made for the two types of target gems in the search task. Virtual gems emit (M = 1917.67, SD = 274.87) lux while physical gems emit (M = 2242.33, SD = 60.02) lux. 

The three physical objects used in the memory task (Figure \ref{fig:recall-objects}) were carefully placed in position before the participants began their trials. These three objects were deliberately chosen to be of different sizes (small, medium, and large), and the three virtual objects used in the memory task were also specifically chosen to have the same range of sizes as the physical ones.


\subsection{Participants}
The study involved 24 adults aged 19 to 32 years (M = 23.88, SD = 4.22), comprising 13 females and 11 males. Nine participants had corrected vision and wore contact lenses during the study, and all were right-handed. Participants were compensated at a rate of \$20 per hour. Their familiarity and experience with VR/AR varied: 20.83\% had no experience with VR, 54.16\% had no experience with AR, 58.33\% had used VR 1-10 times, and 41.66\% had used AR 1-10 times. All procedures were approved by the University's Human Subjects Committee.

\subsection{Procedure}

Participants first signed the informed consent form and completed a pre-study questionnaire collecting demographic information, remembrance of objects, sense of direction (Santa Barbara Sense of Direction Scale; SBSOD~\cite{hegarty2002development}), and experience with AR/VR and gaming. Participants were then fitted with the AR headset and watched a video to orient them to the experiment and explain the search task. They then performed eye gaze calibration within the study headset. Following this, participants completed an eight-minute in-headset tutorial to practice gem classification and spotlight navigation. Only virtual gems were included in the tutorial, though participants were explicitly told they would be searching for both physical and virtual gems in the task. 

The experimenters set up the experimental area out of sight from participants. When the scene was ready, the participant began the trial, which involved walking forwards through the experimental area. Participants were required to only move forwards  so that they could not retrace their steps and revisit an area of the environment multiple times. Once the trial was complete, the participant was taken outside of the experimental area to complete a post-trial questionnaire while an experimenter set up the gems for the next trial. In the post-trial questionnaire, they were asked about the difficulty of the trial they just completed, and whether they saw anything noteworthy in the trial (later used to detect when Godzilla was first noticed). This was repeated for all eight trials. Once all the trials were complete, participants filled out the encountered objects questionnaire to assess memory of objects in the environment, followed by a post-experiment questionnaire for experience ratings. The experiment took approximately 90 minutes in total, including the tutorial and post-experiment survey.

\subsection{Analysis}

Three experimenters reviewed all 192 user trials using a custom playback software to ensure stability and accuracy in user categorization input, and check for misregistrations of responses. Any such misregistrations were corrected by the experimenters before data analysis. Common tactics used by participants included identifying the gem from a distance, categorizing it to the best of their ability, and then re-categorizing it when they had the opportunity to get closer. We did not penalize for re-categorizations, and when a gem had multiple responses, we only considered the latest response for analysis.

We recorded two behavioral metrics during each trial: the total \textit{distance traveled} during the trial (in metres), and the total head rotation measured as the accumulated quaternion distance norm~\cite{huynh2009metrics}. We also measured gem search performance as \textit{detection accuracy} (proportion of gems in the trial that were classified, even if the classification response was incorrect) and \textit{discrimination accuracy} (proportion of detected gems that were correctly classified).

After each trial, participants filled out a post-trial questionnaire measuring \red{cognitive load (as indexed by the }\textit{ease} \red{and \textit{perceived success} in each trial), as well as }their interactions with the spotlight, lighting, and objects. 

\red{Overall trial} \textit{ease} \red{was measured relative to the previous trial (``Was this round easier than the last round?"), and the 7-point Likert scale responses were converted to absolute scores for each trial and normalized} \blue{over the entire dataset.} \red{We also measured the} \textit{\blue{ease}} \red{\textit{of the spotlight} (``How easy was it to follow the spotlight?") for trials that had a spotlight, and \textit{perceived success} in every trial (``In your opinion, how successful were you in identifying the gems?") on a 7-point Likert scale.}

We analyzed the open-ended responses from this questionnaire by implementing grounded theory (GT) techniques, taking a qualitative and inductive approach to derive new conclusions from participant responses~\cite{corbin2014basics}. In this post-trial questionnaire, we asked participants what "noteworthy objects" they found in each trial and read through each response, recording the trends in user feedback. For participants who mentioned noticing the highly salient virtual "Godzilla", we recorded the conditions of the trial in which it was first noticed (low vs. high augmentation density, and path guidance presence). At the conclusion of the study, we verified this information by asking participants to indicate if they had seen the Godzilla, and if so, during which trial they noticed it. We made sure that this information matched the post-trial questionnaire before analyzing it for accuracy and consistency.

Responses to the object memory test were only analyzed for the six objects that were actually present in the environment, out of the twelve presented to participants in the test. The metrics of \textit{recall} ("Was <object> present in the environment") and \textit{precision} (for objects that the user marked present, "Was <object> physical or virtual?") were computed to estimate user memory and spatial awareness. We measured participants' accuracy of \textit{recall} of objects in the environment as the proportion of objects they correctly identified as present/absent, and \textit{precision} as the proportion of recalled (and present) objects that were correctly identified as physical/virtual. 

We conducted factorial analyses within a linear model framework, using the {\small lme4} package in R. For variables that were continuous and appropriately distributed, we used the general linear model. For those that were binomial (gem detection accuracy and gem discrimination accuracy), we used the generalized linear model. For the linear models we report $\beta$, standard errors, 95\% CIs and p-values from Wald chi-square tests.

For the per-trial metrics (distance traveled and total head rotation), the fixed effects were path guidance and augmentation density. The participant ID was included as a random effect. Total head rotation was modeled as a Gaussian distribution, and distance traveled did not satisfy normality and was therefore modeled as a Gamma distribution.

The Godzilla observation data was analyzed using the Chi-square test of independence, and follow-up comparisons for each independent variable (augmentation density and path guidance presence) were performed using the Chi-square goodness of fit test. \red{The cognitive load metrics (}\blue{ease} \red{and perceived success)} \blue{ did not satisfy the normality assumptions of the ANOVA, and were hence analyzed using the ART-ANOVA~\cite{wobbrock2011aligned} from the {\small ARTool} package in R, with results reported as Type III Wald F tests with Kenward-Roger adjusted degrees of freedom. The \textit{spotlight ease} metric also did not satisfy the normality assumption, and was analyzed using the Wilcoxon signed-rank test with the Bonferroni correction for multiple comparisons.} Precision and recall in the object memory test were \blue{also} analyzed using the Wilcoxon signed-rank test from the {\small rstatix} package, with the Bonferroni correction for multiple comparisons.

\begin{figure}[t]
\centering 
\includegraphics[width = \columnwidth]{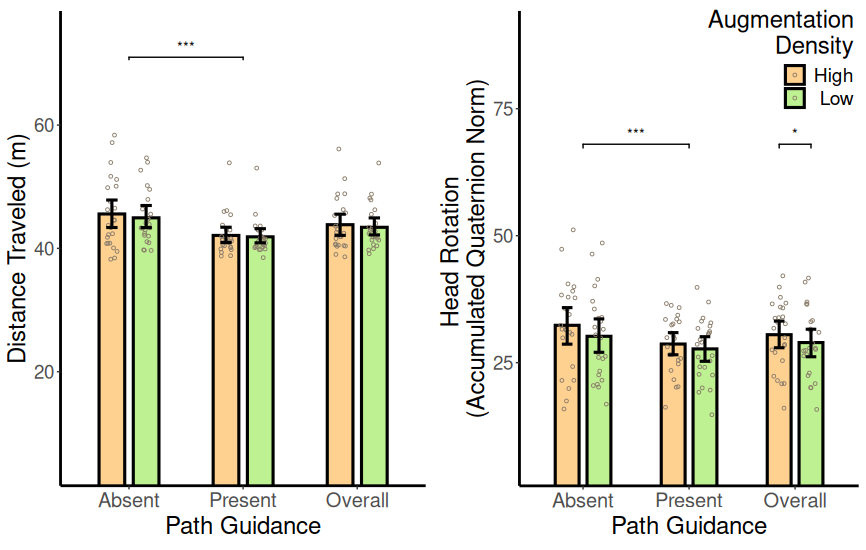}
 \caption{Distance traveled (left) and total head rotation (right), plotted as a function of augmentation density and path guidance. Participants traveled significantly further distances in the absence of the spotlight. Head rotation was significantly higher in trials when the spotlight was absent and was also significantly higher in \red{high augmentation trials vs low augmentation ones}. Gray dots represent data for individual participants and trials. Error bars = 95\% CI.}
 \label{fig:distance-rotation}
 \end{figure}

\section{Results}
The results are organized into four sections. First, we examine participants' physical movement as they navigated through the environment, analyzing distance travelled and head-rotation patterns. Second, we evaluate performance on the gem search task, focusing on gem detection and discrimination accuracy. Third, we assess participants' recall for regular objects encountered in the environment, measured by the number of items recalled in the surprise memory probe at the end of the study. Fourth, we explore recall for the highly salient item, Godzilla, to determine the condition in which it was most likely to be first detected. 

 \begin{figure*}[t]
\centering 
\includegraphics[width = 0.7\textwidth]{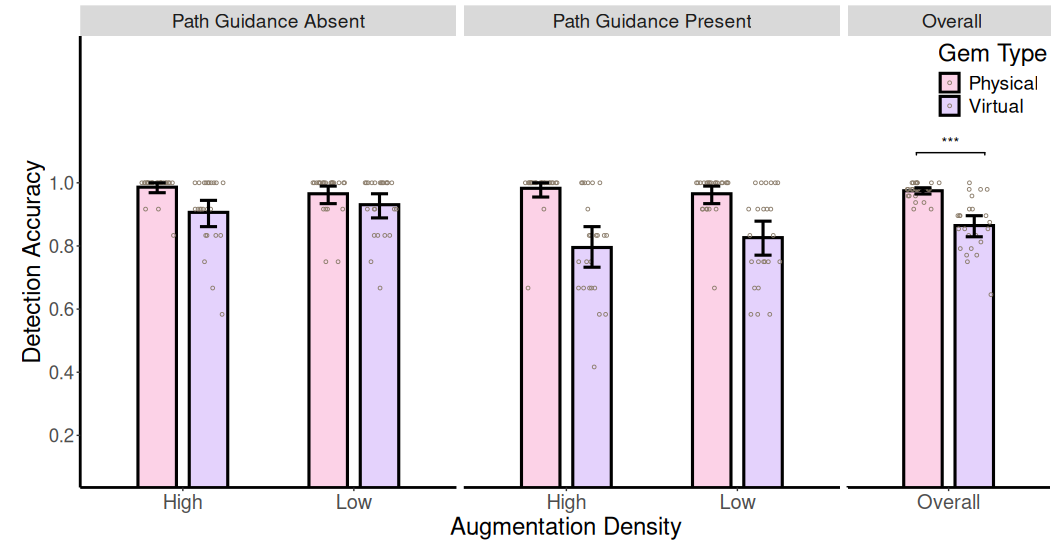}
 \caption{Gem detection accuracy, plotted as a function of augmentation density and gem type for the two path guidance conditions. Overall, participants detected significantly more physical gems than virtual ones. Gray dots represent data for individual participants and trials. Error bars = 95\% CI.}
 \label{fig:detection}
 \end{figure*}

 \begin{figure*}[t]
\centering 
\includegraphics[width = 0.7\textwidth]{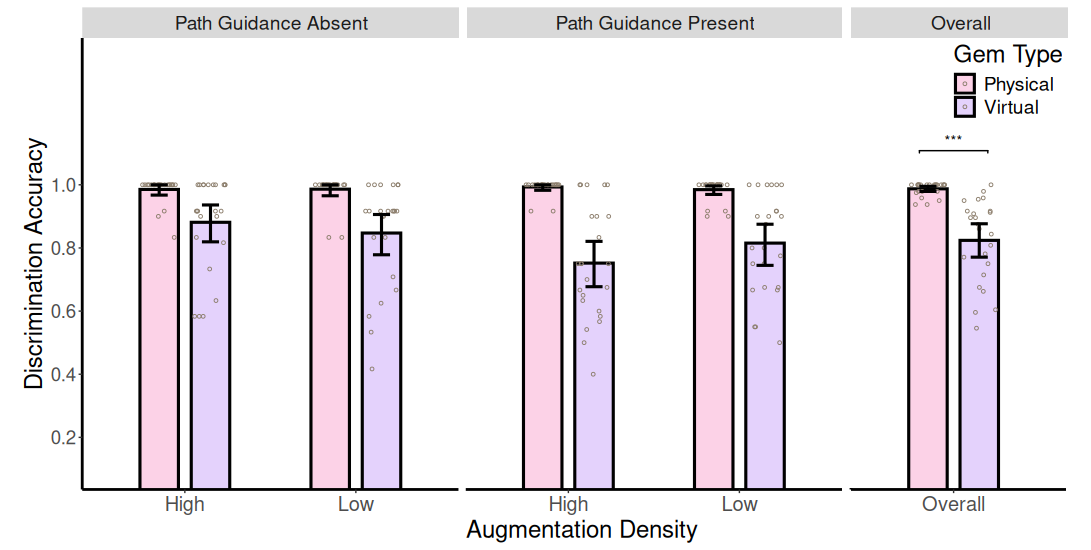}
 \caption{Gem discrimination accuracy, plotted as a function of augmentation density and gem type for the two path guidance conditions. Overall, participants accurately classified physical gems significantly more than virtual ones. Gray dots represent data for individual participants and trials. Error bars = 95\% CI.}
 \label{fig:discrimination}
 \end{figure*}

\subsection{Movement Metrics}
Distance travelled and total head rotation are plotted as a function of augmentation density (low, high) and path guidance (spotlight absent vs. present) in Figure \ref{fig:distance-rotation}.

Augmentation density did not impact distance traveled ($\beta$ = -0.012, $SE$ = 0.012, $p$ $>$ 0.05, $CI$ = (-0.035, 0.011)), but the presence of the spotlight had a significant influence on the distance traveled ($\beta$ = -0.076, $SE$ = 0.012, $p$ $<$ 0.0001, $CI$ = (-0.099, -0.053)), such that participants travelled further when the spotlight was not present. Participants were instructed to maintain their position within the spotlight on spotlight-present trials, but despite these instructions some participants still fell behind on specific trials. These participants and trials are depicted in the outlier data points in Figure \ref{fig:distance-rotation}. There was no significant interaction between augmentation density and path guidance ($\beta$ = 0.007, $SE$ = 0.017, $p$ $>$ 0.05, $CI$ = (-0.025, 0.04)).

Total head rotation was influenced by both augmentation density ($\beta$ = -2.166, $SE$ = 1.05, $p$ $<$ 0.05, $CI$ = (-4.303, -0.021)) and path guidance ($\beta$ = -3.684, $SE$ = 1.05, $p$ $<$ 0.001, $CI$ = (-5.789, -1.569)). If we consider head rotation as a proxy for  scanning of the environment, these data can be interpreted as increased scanning as a function of both conditions of higher augmentation density and also conditions where the spotlight was absent. There was no significant interaction between augmentation density and path guidance ($\beta$ = 1.213, $SE$ = 1.485, $p$ $>$ 0.05, $CI$ = (-1.79, 4.18)).

In summary, distance traveled was influenced by the presence of the spotlight, and head rotation was impacted by both the augmentation density and the presence of the spotlight.

\subsection{Gem Search Performance}
Gem detection accuracy is plotted as a function of augmentation density, gem type and path guidance in Figure \ref{fig:detection}.

Neither augmentation density ($\beta$ = -0.946, $SE$ = 0.593, $p$ $>$ 0.05, $CI$ = (-2.11, 0.216)) nor path guidance  ($\beta$ = -0.228, $SE$ = 0.67, $p$ $>$ 0.05, $CI$ = (-1.541, 1.085)) impacted gem detection accuracy, but gem type was a significant predictor of detection accuracy ($\beta$ = -2.024, $SE$ = 0.539, $p$ $<$ 0.001, $CI$ = (-3.08, -0.969)), with physical gems being detected significantly more than virtual gems. There were no significant two-way or three-way interactions (all $\beta$ $<$ 1.281, $SE$ $<$ 0.892, $p$ > 0.05).

Gem discrimination accuracy is plotted as a function of augmentation density, gem type and path guidance in Figure \ref{fig:discrimination}. Gem discrimination results were similar to gem detection, such that neither augmentation density ($\beta$ = -0.021, $SE$ = 0.706, $p$ $>$ 0.05, $CI$ = (-1.404, 1.362)) nor path guidance  ($\beta$ = 0.694, $SE$ = 0.861, $p$ $>$ 0.05, $CI$ = (-0.994, 2.381)) impacted gem discrimination accuracy, but gem type was a significant predictor of detection accuracy ($\beta$ = -2.342, $SE$ = 0.536, $p$ $<$ 0.0001, $CI$ = (-3.393, -1.29)), with physical gems being correctly discriminated significantly more than virtual gems. There were no significant two-way or three-way interactions (all $\beta$ $<$ 1.373, $SE$ $<$ 1.168, $p$ > 0.05).

In summary, both gem detection and discrimination were both only significantly influenced by the gem type, with higher accuracies for the physical gems as compared to virtual ones.

\subsection{\red{Cognitive Load}}

 \begin{figure}[t]
\centering 
\includegraphics[width = \columnwidth]{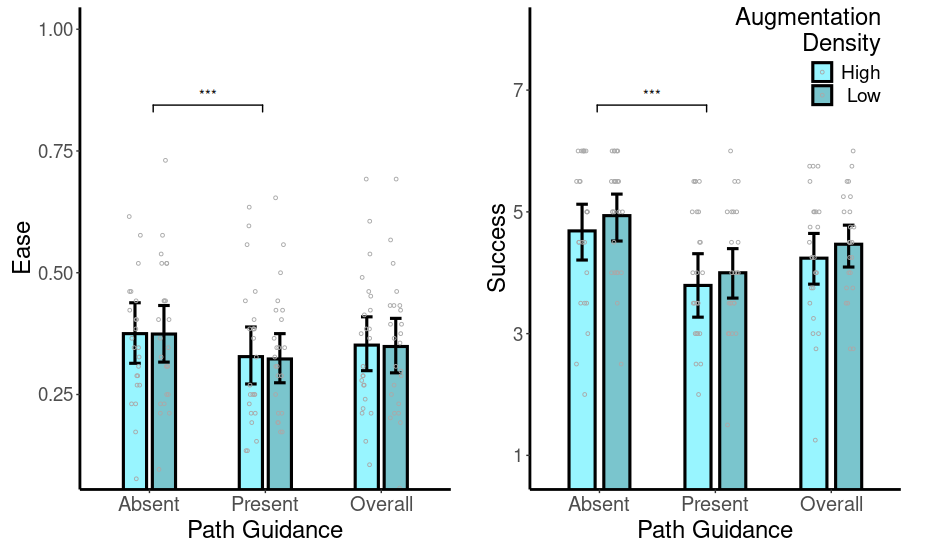}
 \caption{\red{Overall} \blue{ease} \red{ (left) and perceived success (right) as reported in the post-trial questionnaire, higher scores indicate higher} \blue{ease} \red{and success respectively. Gray dots represent data for individual participants. Error bars = 95\% CI.}}
 \label{fig:cog-load}
 \end{figure}

\blue{ There was a main effect of path guidance on the self-reported ease of the trial, with lower ease scores when the path guidance was present (F(1, 165) = 18.165, $p$ $<$ 0.0001). There was no main effect of augmentation density (F(1, 165) = 0.203, $p$ $>$ 0.05) or interaction (F(1, 165) = 0.013, $p$ $>$ 0.05).

There was no effect of augmentation density on the ease of following the spotlight ($V$ = 153, $p$ $>$ 0.05, $r$ =.282, $small$).

There was a main effect of path guidance on the perceived success in the task, with lower perceived success when the path guidance was present (F(1, 165) = 35.006, $p$ $<$ 0.0001). There was no main effect of augmentation density (F(1, 165) = 0.54, $p$ $>$ 0.05) or interaction (F(1, 165) = 0.079, $p$ $>$ 0.05).
}

\red{In summary, the presence of the path guidance led to} \blue{lower ease} \red{and lower perceived success in the task. Augmentation density did not impact any of our metrics of cognitive load.
}

\subsection{Memory for Goal-Irrelevant Objects in the Environment}
 
\begin{figure}[t]
\centering 
\includegraphics[width = \columnwidth]{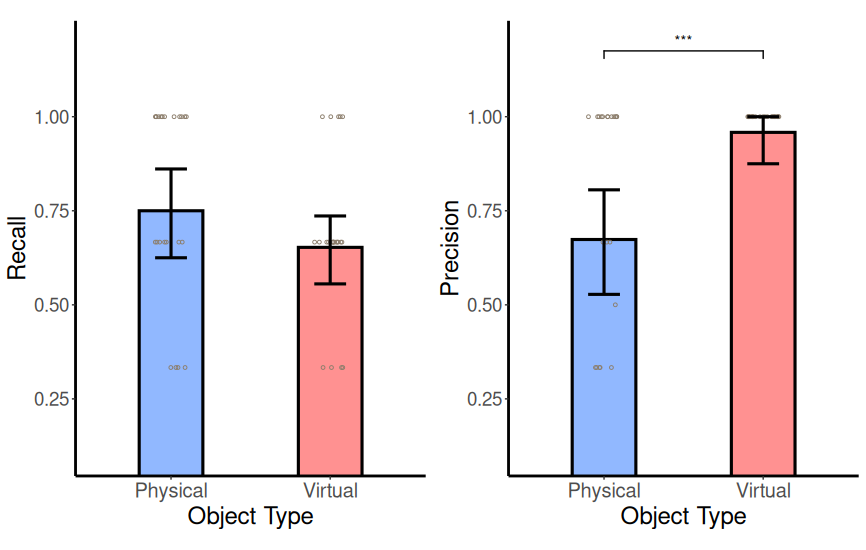}
 \caption{Recall (left) and precision (right) on the post-experiment object memory test, for objects that were present in the environment. 1 represents a perfect score for both metrics, and 0 the lowest score. Black dots represent data for individual participants. Error bars = 95\% CI.}
 \label{fig:recall}
 \end{figure}

There was no significant difference in the recall of physical vs. virtual objects ($V$ = 69.5, $p$ $>$ 0.05, $r$ = .288, $small$), as shown in Figure \ref{fig:recall}. However, recall precision (i.e. the proportion of recalled objects that were correctly classified as physical or virtual) was significantly higher for virtual objects as compared to physical objects ($V$ = 2.5, $p$ $<$ 0.01, $r$ = .605, $large$). Thus, although there does not appear to be a difference in recall of physical vs. virtual objects, participants experienced source confusion much more with physical objects as compared to virtual ones.

\subsection{Awareness of Highly Salient Item (Godzilla)}

Here, we examined whether the level of augmentation density and presence or absence of path guidance influenced participants' ability to detect and remember the virtual Godzilla figure. Given that Godzilla was constantly present in the environment, we were able to obtain two metrics. First, did participants detect Godzilla at all during the experiment? Second, if detected, under what conditions (augmentation density and path guidance) did detection first occur?

The task was relatively challenging but still feasible, as 16/24 (66.66\%) of participants noticed Godzilla. A Chi-square test of independence indicated that the experimental condition had an impact on awareness of Godzilla ($\chi^{2}$(1, $N$=24) = 15, $p$ $<$ 0.0005). Participants noticed Godzilla significantly more in the absence of path guidance ($\chi^{2}$(1, $N$=24) = 61.25, $p$ $<$ 0.0005), with only one participant noticing it during a spotlight-present condition. The augmentation density also influenced observation of Godzilla, with participants noticing it significantly more in the low-augmentation condition ($\chi^{2}$(1, $N$=24) = 5.33, $p$ $<$ 0.025). Furthermore, a Mann-Whitney U test showed that head rotation was significantly higher in participants who noticed Godzilla ($U$ = 19, $Z$ = -2.84, $p$ $<$ 0.005, $\eta_{}^{\mathrm{2}}$ =1.17), suggesting that Godzilla detection likelihood was greater in participants who scanned the environment more extensively. 

\subsection{Subjective Participant Feedback}
In this section, we analyze and discuss the qualitative feedback provided by participants, highlighting key themes that emerged from their experiences during the trials. 

Participants did notice the difference between lightly and heavily augmented trials, with one participant {(P6)} noting, "In heavily augmented trials, it was harder to identify gems and objects sometimes.'' Interestingly, participants also commented that ``With background trial (high-augmentation), it was easier for me to find physical gems" {(P13)}, indicating a potential divide between the physical and virtual environments which made it easier to focus on the physical environment when the virtual environment was cluttered.

This divide between the virtual and physical worlds was further evidenced by the results of the gem search task. Participants detected physical targets at a significantly higher rate than virtual targets across all conditions, even though both types of targets were designed to be as visually identical as possible, with similar relative brightness to the surrounding objects. This difference in detection could be attributed to an increased focus on the physical environment over the virtual, though  the limited field of view of the virtual environment compared to the physical one may have also contributed to this effect.

Following the spotlight was designed to mimic regular walking, as the guided pace matched a natural walking speed during the search task. One participant {(P11)} mentioned, "Following the ring felt similar to regular walking, as the guided pace was consistent with a natural walking speed.'' While some participants felt the spotlight was too fast, others found it slower than their usual pace, which allowed them to spend more time on task. Participant {(P24)} stated, "The spotlight's slower pace in one of the trials allowed me to spend more time looking closely at the gems.'' Ultimately, each individual's comfort level with these spatial UI interfaces and their typical task performance speed played a significant role, though most participants managed to keep up with the designated timing. This precision is optimal for our goal of studying attention effects in a dynamic, “on the go” manner by enforcing controlled walking motions. Our approach involves balancing the benefits of a predetermined and enforced walking path with those of a predetermined but more flexibly followed walking path.

This feedback provided valuable insights that should inform the design considerations for AR applications intended for use on the go. Considerations include managing the balance between physical and virtual elements and designing guided navigation that aligns with natural walking speeds while accommodating individual user preferences. We elaborate on these in the following section.

\section{Discussion}

AR technology allows us to integrate virtual content into the physical world and has potential to be a primary source of information consumption on the go. While these technologies hold much promise, it is crucial to address the potential issues associated with AR clutter. This visual overload has the potential to distract users from their primary tasks, and perhaps even jeopardize their safety by reducing awareness of the physical surroundings.  To address these potential concerns, we designed a study to test the impact of varying levels of augmentation and path guidance on participants' performance  and awareness of objects in their environment in a locomotion-based search task.  

Our first goal was to examine the impact of varying levels of augmentation on user behavior and awareness of their environment. While there did not appear to be any change in search task performance (gem detection and discrimination accuracy) with augmentation density, participants did look around the environment significantly more (as evidenced by increased head rotation) in the high augmentation condition. This suggests that participants opted to scan the environment much more to find gems when augmentation density was high, which was supported by participant comments in the post-study interview. Similarly, participants spotted Godzilla significantly less in the high augmentation trials. These results suggest that increased levels of augmentation clutter in the environment can exacerbate inattention blindness and make exploring the environment more challenging. 

There was no difference in the recall of physical and virtual objects from the post-study object recall test, though participants falsely remembered some physical objects as virtual but no virtual objects as physical, a source confusion phenomenon previously reported in the literature \cite{fernandes2015remembering}. One contributing factor to this source confusion effect might be the novelty effect of the virtual augmentations. Notably, 19 out of 24 participants had prior VR headset experience, while 13 out of 24 had never tried head-mounted AR devices such as the HoloLens 2. 

Previous research involving unannounced recall tasks \cite{kumaran2023impact, kim2022investigating,  tatzgern2016adaptive} has shown that participants performing a search task with only virtual targets had a much better awareness of the virtual environment compared to the physical one. The inclusion of physical targets in our experiment appears to reduce this effect, suggesting that the nature of the search targets impacts awareness of different aspects of the environment. \red{This is good news to mobile AR designers to some extent, as the previous results left open the possibility that use of AR {\em generally} diminishes recall of real-world objects and infrastructure if not carefully designed. }
 
 Our results also further support the hypothesis that there is a competing focus effect of the virtual and physical environments, and participants tend to focus on one or the other depending on the nature of the task. In some of the prior work\cite{kim2022investigating, kumaran2023impact}, the task only required attention to the virtual environment and attention to the physical environment was important primarily for safe navigation, whereas our experiment demanded attention to both virtual and physical spaces. And the perceived nature of the task at hand (whether one is operating in virtual space or in physical space) definitely matters. A study comparing three-week-long care for a virtual pet in AR and VR~\cite{poretski2019virtual} for example led to vastly different feelings of ownership and emotional connection. Surely, task virtuality can influence focus and recall of objects as well, such as to result in higher recall of virtual objects when the task is perceived to be a primarily virtual task. 

In our case, the lack of difference in recall for the two types of objects, while obviously not a finding in itself, is noteworthy when compared to the mentioned previous research and seems to indicate that our task is indeed perceived as taking place in both the virtual and physical realm. \red{With our 50-50 split of virtual and physical gems, the search results suggest higher participant attention to the physical environment, as evidenced by the higher detection accuracy for physical gems.} At the same time, the novelty of the virtual objects could have been a competing factor that led to the observed results in the recall test, with higher source confusion for physical objects. 

Participants were also worse at discriminating surface markings for the virtual gems compared to physical ones, which was likely due to the HoloLens 2's display aliasing which made the markings on virtual gems harder to discern from a distance. This issue was not observed in our pilot studies, perhaps because pilot participants were slightly more expert with AR and more familiar with the optical see-through display technology. 

The presence of the path guidance spotlight altered movement behavior, with reduced distance traveled, reduced head rotation, and strongly reduced likelihood of spotting the highly conspicuous Godzilla model. \red{The spotlight also} \blue{decreased both self-reported ease and perceived success} \red{in the task, }although most participants managed to stay within the designated timing, demonstrating that the interface effectively guides users with precise timing. Additionally, the presence of the spotlight did not impact the task performance metrics (which were only affected by the gem type). 

\red{For appreciating the broader significance of our results, imagine a future where people routinely utilize lightweight AR glasses while walking, either in unfamiliar environments, where a use case might be navigation and wayfinding, or during daily common familiar walks, where a use case might be parallel consumption of AR for such purposes as entertainment, education, communication, corporate duties, or personal information management. Our results demonstrate that AR tools, even if effective at the function they are designed for, may have negative side effects. Application designers must carefully consider the benefits and costs of AR annotations and tools, and focus on minimizing adverse effects on user experience. }
For our path guidance tool, participants' subjective feedback suggested that the level of comfort with the spotlight varied depending on each users' general walking pace, indicating a potential advantage of personalization for guidance tools of this kind.

\subsection{Design Considerations}
Here, we share our guidance and recommendations for the design of on-the-go AR applications, based on insights gathered from our physical and virtual natural locomotion study.

We found that increased clutter in environments can exacerbate inattention blindness. Therefore, designers should strive to reduce unnecessary clutter and highlight or draw attention to important objects, both task-relevant and task-irrelevant, especially when these objects are crucial for user safety. 

The nature of a user's task focus--whether primarily physical or virtual--affects their awareness of various environmental aspects. For tasks that require focus on the virtual environment, designers should find ways to emphasize important physical objects to ensure safe navigation in the real world.

Effective mechanisms to control walking paths can ensure safe navigation without impairing the user's focus on their main task. Personalizing these tools based on the user's walking behavior and familiarity with different spatial interfaces can enhance their effectiveness. Additionally, matching the task with the user's natural walking speed can provide a seamless experience, allowing users to concentrate on the task while being safely guided, without compromising performance.

\subsection{Limitations}

While all our efforts were focused on ensuring experimental rigor and controlling unwanted impact factors and dependencies, our study does have limitations. First, the HoloLens 2 has a relatively small field of view as compared to some newer AR HMDs. We are familiar with the existing literature on FOV limitations and their potential impacts, as discussed in Section \ref{sec:related-work}. However, the HoloLens 2 was the most suitable device for this task due to the need for an optical-see-through headset with stable tracking and registration in larger environments (which we achieved with MRTK's world locking tools). When advancements in FOV and resolution are achieved with the next generation of AR devices, follow-up studies should be conducted to determine if these effects are still observed. 

Second, we put a lot of effort into trying to ensure a similar appearance of physical and virtual gems, but some challenges, such as uneven color appearance at different visual angles in the HoloLens 2 display could not be addressed by our modeling and rendering efforts. Positive comments and reactions from some participants on their first encounter of both physical and virtual targets give us confidence that our efforts were worthwhile, but in the end, OST-AR technology does not currently support a truly seamless integration of real and virtual space, and it remains to be seen if some effects would come out slightly different with more accurate visual coherence.   

Third, while our intention for this study was to inform realistic future AR-on-the-go scenarios, the need to run a controlled user study meant that we needed to pick a particular representative task. Our choice fell on target search as a well-studied attention task that is a likely element of any AR scenario involving the physical space around a mobile user, but there will clearly be some other possible AR-on-the-go tasks for which this choice represents but limited ecological validity. Still, we believe that a sizable set of future AR efforts can benefit from the initial results presented here. 

Fourth, there are still alternative hypotheses regarding the lack of difference in recall between physical and virtual objects, when compared to previous work that found diminished recall of physical objects. We believe that the introduction of physical targets in our study, balanced with virtual targets, is the main factor behind the absence of differing recall accuracy, but our experiment had also some other key differences with the work we compared results to, namely a smaller search environment and being indoors vs. outdoors. It is possible that these differences may have played a role too, and future experiments will be needed to examine these additional factors. 

\section{Conclusion}
Through this project, we examined how varying levels of augmentation affect user behavior and awareness of their environment. Our findings show that high augmentation density leads to increased head rotation and environmental scanning, suggesting that participants needed to search more thoroughly to find virtual gems. User comments supported this, noting difficulties in identifying gems amidst heavy augmentations and the relative ease of finding physical gems in cluttered virtual settings (an indication that there is clearly still a divided-worlds effect -- clutter in virtual annotation space does not necessarily impact the physical world to the same extent). The higher detection rate of physical targets over virtual ones, despite their visual similarities and focus on virtual targets in the tutorial, further emphasizes divide between the physical and virtual world. The limited virtual-environment field of view of the AR headset most likely plays a role here. 

In a post-study recall test, there was no significant difference in recalling physical versus virtual objects, although participants sometimes misremembered physical objects as virtual, a previously demonstrated source-confusion effect likely influenced by the novelty of virtual augmentations. Previous studies showed higher recall of virtual objects when tasks are focused solely on the virtual environment. The nature of the task at hand -- is it a task in the virtual realm, in the physical realm, or in both? -- appears to make a big difference concerning scene and object awareness.

This study highlights the complex interplay between physical and virtual environments in augmented reality, emphasizing the need for better AR design to integrate virtual elements seamlessly into the physical world and enhance user engagement \red{while walking. It provides insights about the impact of AR content on walking users' behavior and cognitive processes, which is relevant for such tasks as navigation and wayfinding in unfamiliar environments, or for personal applications (entertainment, education, personal information management) that make use of referencing the physical world during common daily walks.}

\begin{acks}
This work was funded by NSF Award IIS-2211784, the Army Research Office under contract W911NF-19-D-0001, the Office of Naval Research with grant N00014-23-1-2118, and DARPA under contract HR001124C0409.
\end{acks}



\bibliographystyle{ACM-Reference-Format}
\bibliography{sample-base}


\end{document}